\documentclass[slac_one]{revtex4}
\usepackage{graphicx}
\usepackage{fancyhdr}
\begin{document}

\title{Tau trigger at the ATLAS experiment} 

\author{K.~Benslama}
\affiliation{University of Regina, Regina, Canada}

\author{C.~Belanger-Champagne}
\affiliation{University of Uppsala, Uppsala, Sweeden}

\author{M.~Bosman}
\affiliation{The Institute for High Energy Physics (IFAE),  Barcelona, Spain}

\author{R.~Brenner}
\affiliation{University of Uppsala, Uppsala, Sweeden}

\author{P.~Casado}
\affiliation{The Institute for High Energy Physics (IFAE), Barcelona, Spain}

\author{Z.~Czyczula}
\affiliation{University of Copenhagen, Copenhagen, Denmark}

\author{M.~Dam}
\affiliation{University of Copenhagen, Copenhagen,  Denmark}

\author{S.~Demers}
\affiliation{Stanford Linear Accelerator Center (SLAC), USA}

\author{S.~Farrington}
\affiliation{University of Oxford, Oxford, UK}

\author{O.~Igonkina}
\affiliation{NIKHEF, Amsterdam, The Netherlands}

\author{A.~Kalinowski \footnote{Corresponding author} } 
\affiliation{University of Regina, Regina, Canada}

\author{N.~Kanaya}
\affiliation{University of Tokyo, Tokyo, Japan}

\author{C.~Osuna}
\affiliation{The Institute for High Energy Physics (IFAE), Barcelona, Spain}

\author{E.~Perez}
\affiliation{The Institute for High Energy Physics (IFAE),  Barcelona, Spain}

\author{E.~Ptacek}
\affiliation{University of Oregon, Eugene, USA}

\author{A.~Reinsch}
\affiliation{University of Oregon, Eugene, USA}

\author{A.~Saavedra}
\affiliation{University of Sydney, Sydney, Australia}

\author{A.~Sfyrla}
\affiliation{University of Illinois, Urbana-Champaign, USA}

\author{A.~Sopczak}
\affiliation{Lancaster University, Lancaster, UK}

\author{D.~Strom}
\affiliation{University of Oregon, Eugene, USA}

\author{E.~Torrence}
\affiliation{University of Oregon, Eugene, USA}

\author{S.~Tsuno}
\affiliation{University of Tokyo, Tokyo, Japan}

\author{V.~Vorwerk}
\affiliation{The Institute for High Energy Physics (IFAE), Barcelona, Spain}

\author{A.~Watson}
\affiliation{University of Birmingham, Birmingham, UK}

\author{S.~Xella}
\affiliation{University of Copenhagen,  Copenhagen, Denmark}

\begin{abstract}
Many theoretical models, like the Standard Model or SUSY at large tan($\beta$), predict Higgs bosons or new particles which
decay more abundantly to final states including tau leptons than to other leptons. At the energy scale of the LHC, the
identification of tau leptons, in particular in the hadronic decay mode, will be a challenging task due to an overwhelming
QCD background which gives rise to jets of particles that can be hard to distinguish from hadronic tau decays. Equipped with
excellent tracking and calorimetry, the ATLAS experiment has developed tau
identification tools capable of working at the trigger level. This contribution presents tau trigger algorithms which exploit
the main features of hadronic tau decays and describes the current tau trigger commissioning activities.
\end{abstract}

\maketitle

\thispagestyle{fancy}

\section{THE ATLAS TAU TRIGGER}
The ATLAS trigger system is divided into two main parts: the Level 1 (LVL1)~\cite{ L1-TDR}, hardware-based trigger, and the High
Level Trigger (HLT)~\cite{HLT-TDR}, using software selection algorithms. The hardware implementation of the LVL1 trigger is 
determined by very strong constraints in the processing time due to the high bunch-crossing rate of 40 MHz. Only a few $\mathrm{\mu s}$
are available for the decision making at this level. Due to the rate reduction achieved at LVL1 more decision time is allowed in the two
stages of selection at the HLT: a few tens of milliseconds at the Level 2 Trigger (LVL2), and a few seconds at the Event Filter
(EF). In this section a brief description of the tau trigger will be presented.
\subsection{Level 1 tau trigger}
\begin{figure*}[!h]
\centering
\includegraphics[width=45mm]{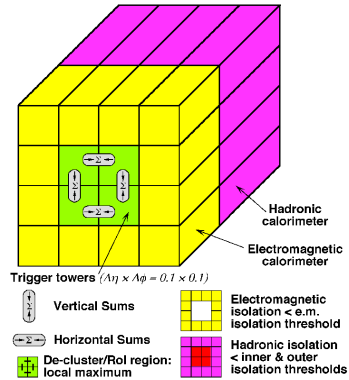}
\caption{Trigger towers used in the level 1 tau trigger.} \label{L1Scheme}
\end{figure*}

The LVL1 trigger~\cite{L1-Calo} uses trigger towers $\Delta \eta \times \Delta \phi$=0.1$\times$0.1
 within the electromagnetic and hadronic calorimeter to identify Regions of Interest (RoI) 
to be investigated further in the HLT. The following quantities are used to select LVL1
features giving rise to a RoI, see Fig.~\ref{L1Scheme}: 
a) energy in 2$\times$1 pairs of EM towers within a 2$\times$2 core region; 
b) energy in a 2$\times$2 group of hadronic towers behind the EM core; 
c) energy in a 12 tower EM isolation ring surrounding the central 2$\times$2 core; 
d) energy in a similar ring in the Hadronic calorimeter.  
Relatively low thresholds are applied at LVL1 and are subsequently refined
at the HLT. Triggers can be defined with and without an isolation requirement.
\subsection{High Level Trigger}
Due to the greater time available for decisions at the HLT, the cluster information can be refined using the full granularity of
the calorimeter in conjunction with a cell energy calibration and tracking information from the inner detector 
can be included~\cite{ATLAS-detector}.
The characteristics used to distinguish tau decays from particle jets are:
\begin{figure}[!htbp]
  \begin{center}
    \begin{tabular}{ccc}
    \resizebox{0.35\linewidth}{!}{\includegraphics{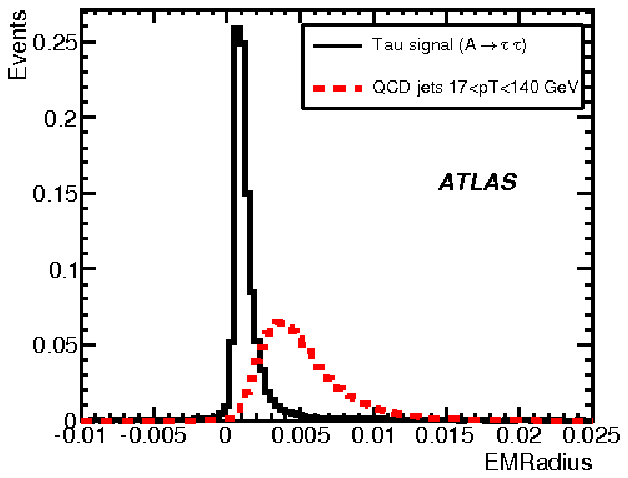}} & \hspace{0.1\linewidth} &
    \resizebox{0.2\linewidth}{!}{\includegraphics{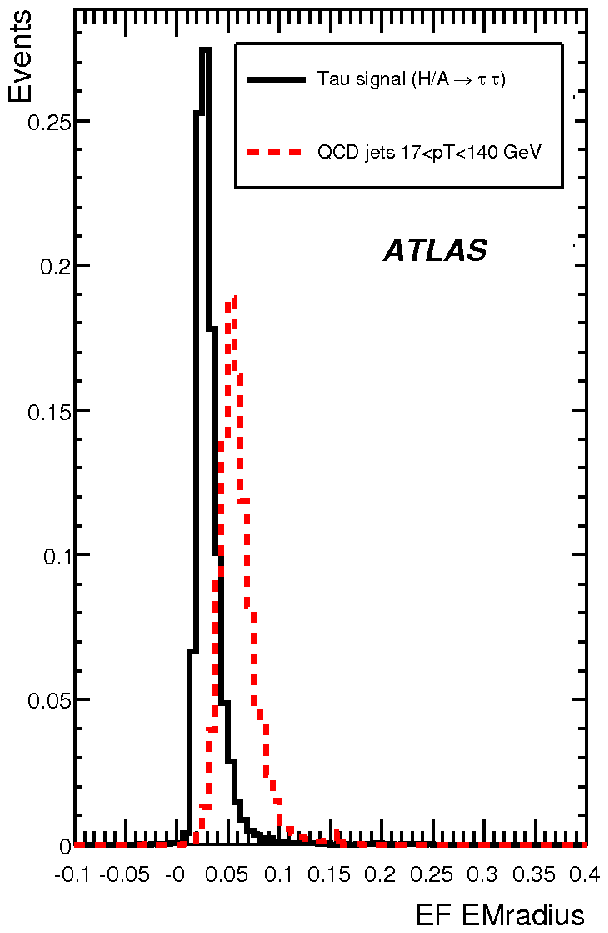}}
   \end{tabular}
  \caption{Distribution of the energy-weighted mean squared cluster radius in the EM calorimeter
           Left plot shows the quantity calculated at the LVL2, 
           and right plot presents the same quantity calculated at the EF. The taus from $\mathrm{A\rightarrow \tau \tau}$ (with
           $\mathrm{ m_{A}=800~GeV/c^{2}}$) were used as signal events, and QCD dijets with hard scattering $\mathrm{p_{T}}$ in range 
           of $\mathrm{17-140~GeV}$ were used as background events.
            \label{EMRadius}}       
  \end{center}
 \end{figure}
low charged track multiplicity, calorimeter cluster isolation and narrowness of the jet.
The narrowness can be estimated using several different variables, Figure~\ref{EMRadius} shows the electromagnetic radius (EMRadius), 
which is the energy-weighted mean squared cluster radius in the EM calorimeter. Figure~\ref{EMRadius} (left) shows the 
EMRadius calculated at LVL2, where
only information from a limited region around the LVL1 cluster is used, and Figure~\ref{EMRadius} (right) shows the same
quantity calculated at EF, where the information from the whole detector is available and slower and more precise algorithms can
be used. Typical rejections against jets of the order of 20 are achieved at the HLT (LVL2 + EF).

The HLT tau slice algorithms have been commissioned within the ATLAS online environment at the LHC Point 1. Here the full
execution time, including data access and unpacking can be measured. Routine online tests using simulated data and
a complete trigger menu, which includes a full set of tau signatures, show that the
timing meets the online requirements.
\subsection{Tau trigger online monitoring}
\begin{figure}[!htbp]
  \begin{center}
    \begin{tabular}{ccc}
  \resizebox{0.35\linewidth}{!}{\includegraphics{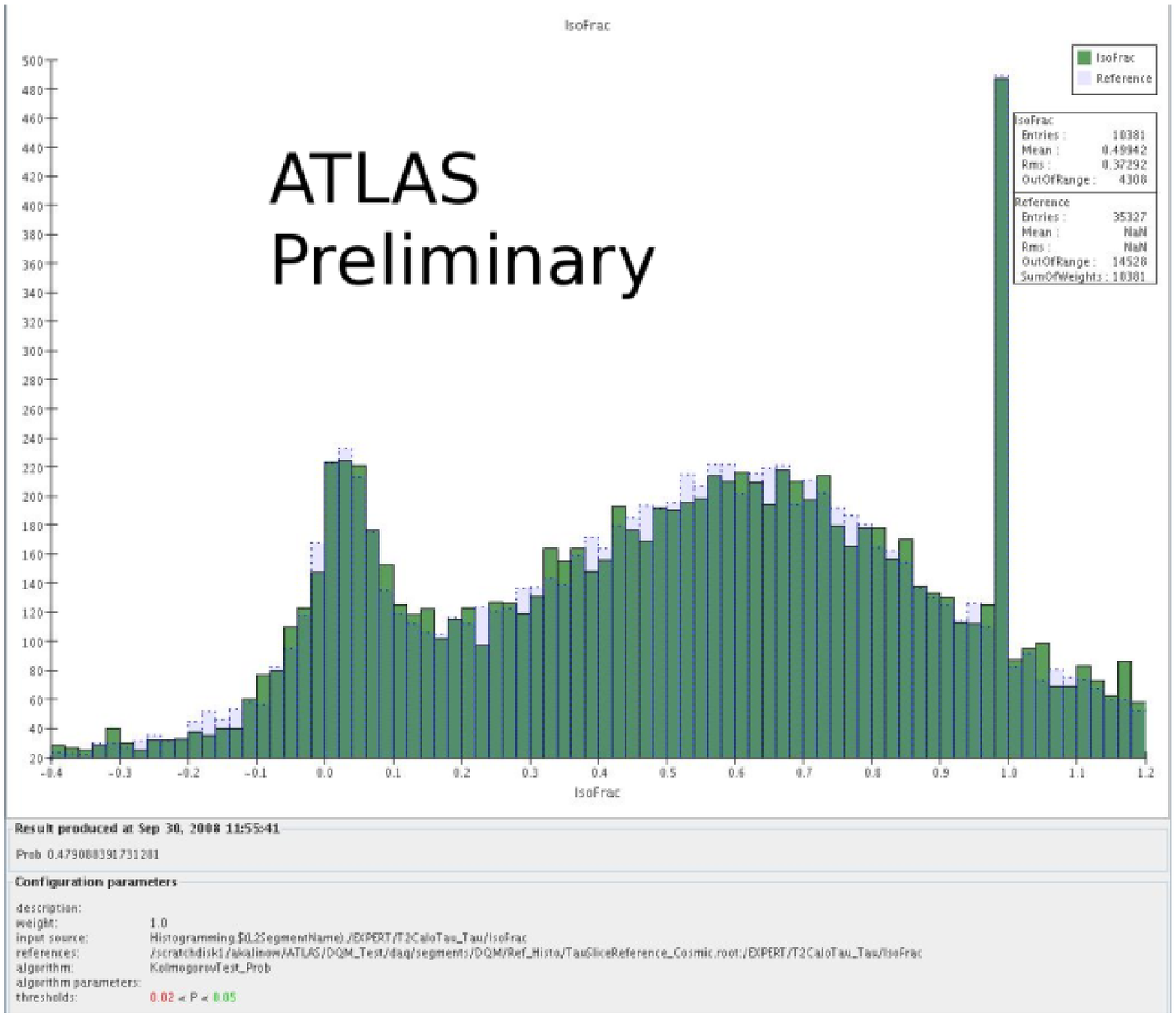}} & \hspace{0.1\linewidth} &
  \resizebox{0.35\linewidth}{!}{\includegraphics{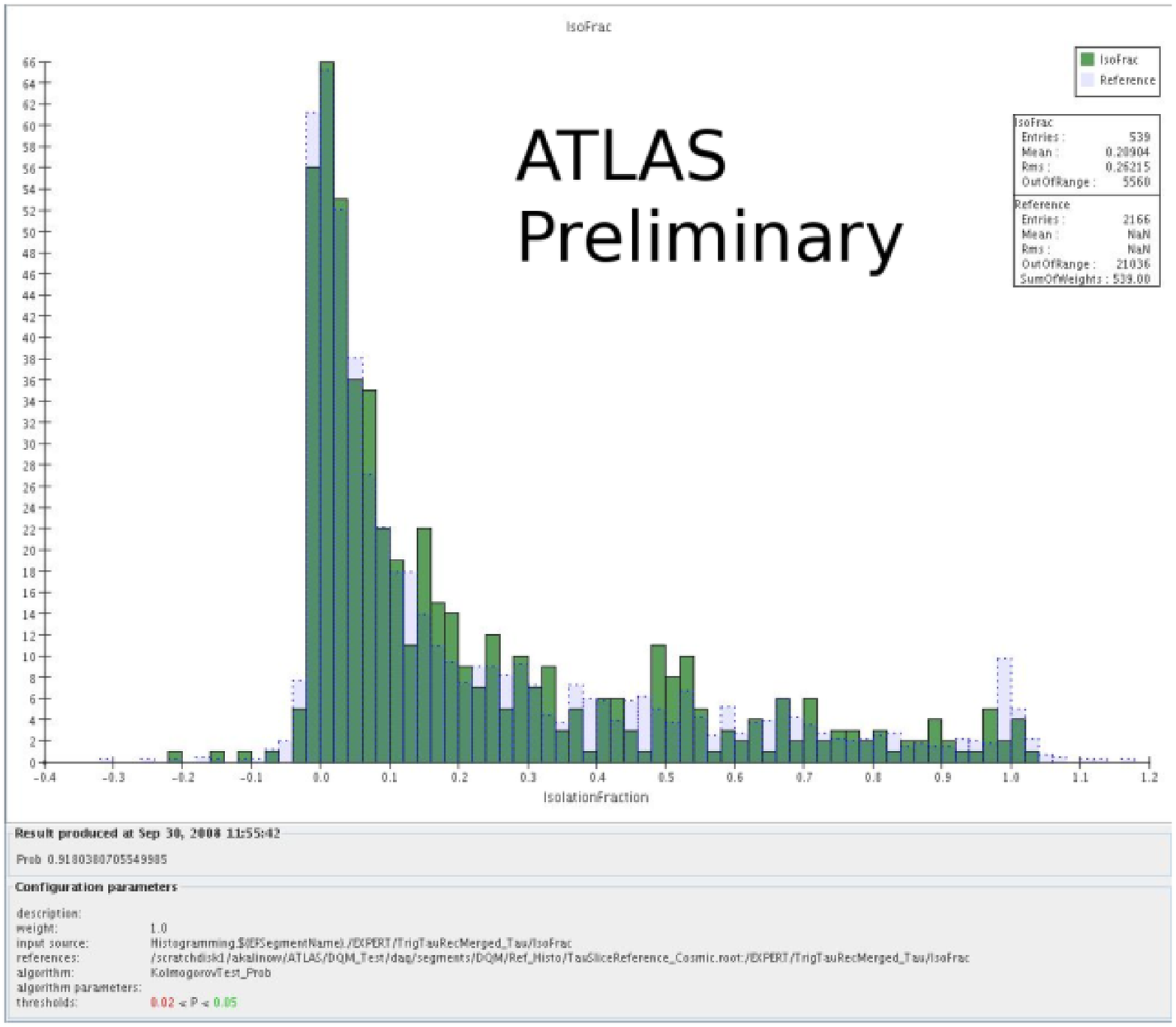}} 
   \end{tabular}
  \caption{Screen-shot of the online data quality monitoring display application. Left plot presents the isolation fraction
           variable monitored at the LVL2, and right plot presents the same variable monitored at the EF. The histograms are 
           made for the ATLAS runs with cosmic muons.
            \label{DQMF}}       
  \end{center}
 \end{figure}

The trigger performance will be monitored online with the ATLAS Data Quality Monitoring Framework. 
All the variables used to identify the tau candidates will be histogrammed, and an automatic check 
of the histogram validity will be performed. The tests include fitting known distribution shapes 
and comparison with the reference histogram.
Figure~\ref{DQMF} presents a screen-shot of the monitoring application, showing one of the monitoring histograms. The system
is already in use at the ATLAS online environment, and has been tested with the first single beam LHC data.

In early running the performance of the tau trigger will be determined from data using events that feature QCD interactions
where some jets closely mimic tau leptons and from a sample of  $\mathrm{Z\rightarrow \tau \tau}$ decays. 
In the latter case, a tag \& probe method will be used to measure the tau trigger efficiency using a clean sample of events 
containing hadronically decayed $\tau$ leptons. 
\begin{figure}[!htb]
  \begin{center}
    \begin{tabular}{ccc}
  \resizebox{0.35\linewidth}{!}{\includegraphics{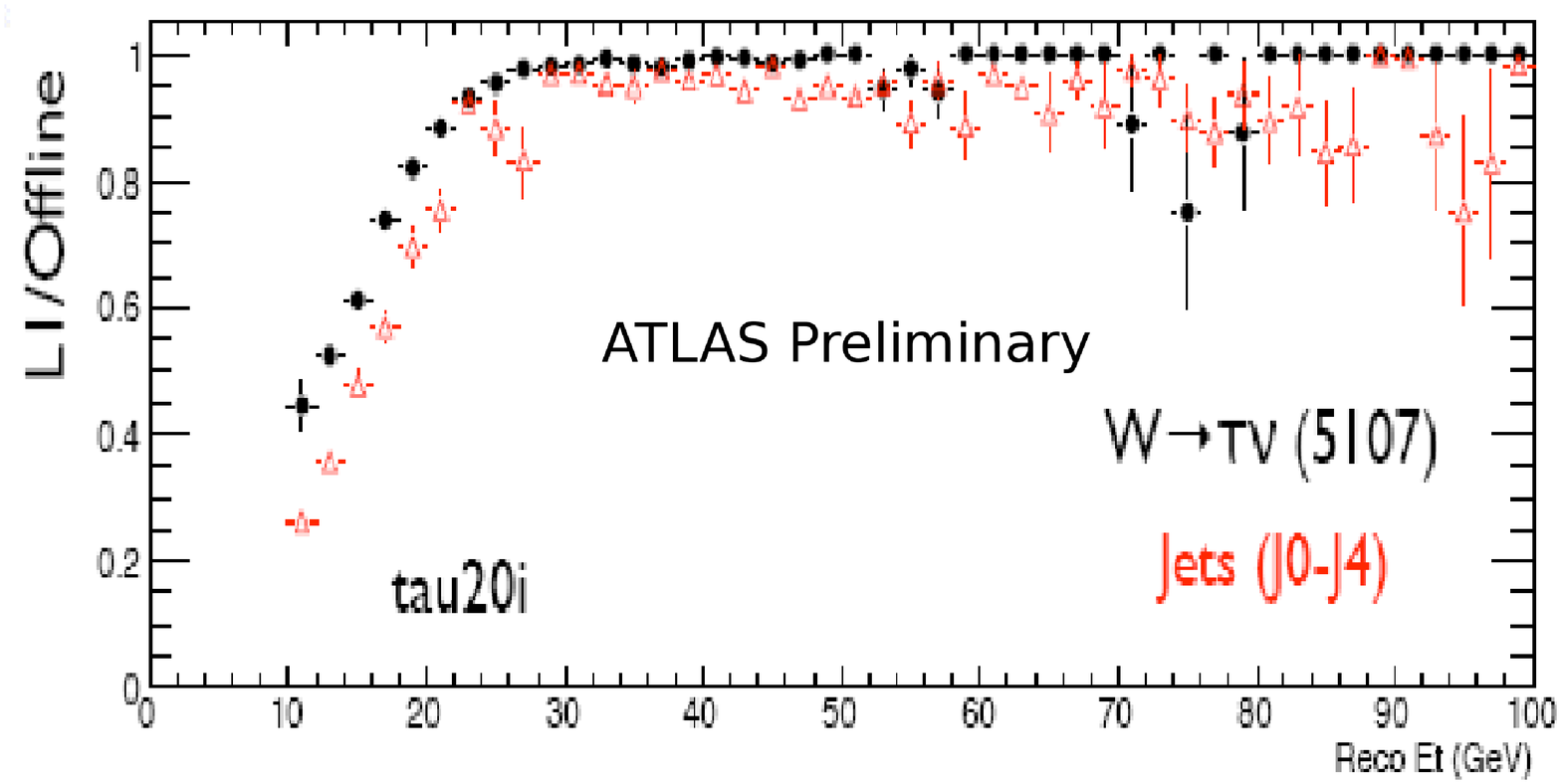}} & \hspace{0.1\linewidth} &
  \resizebox{0.3\linewidth}{!}{\includegraphics{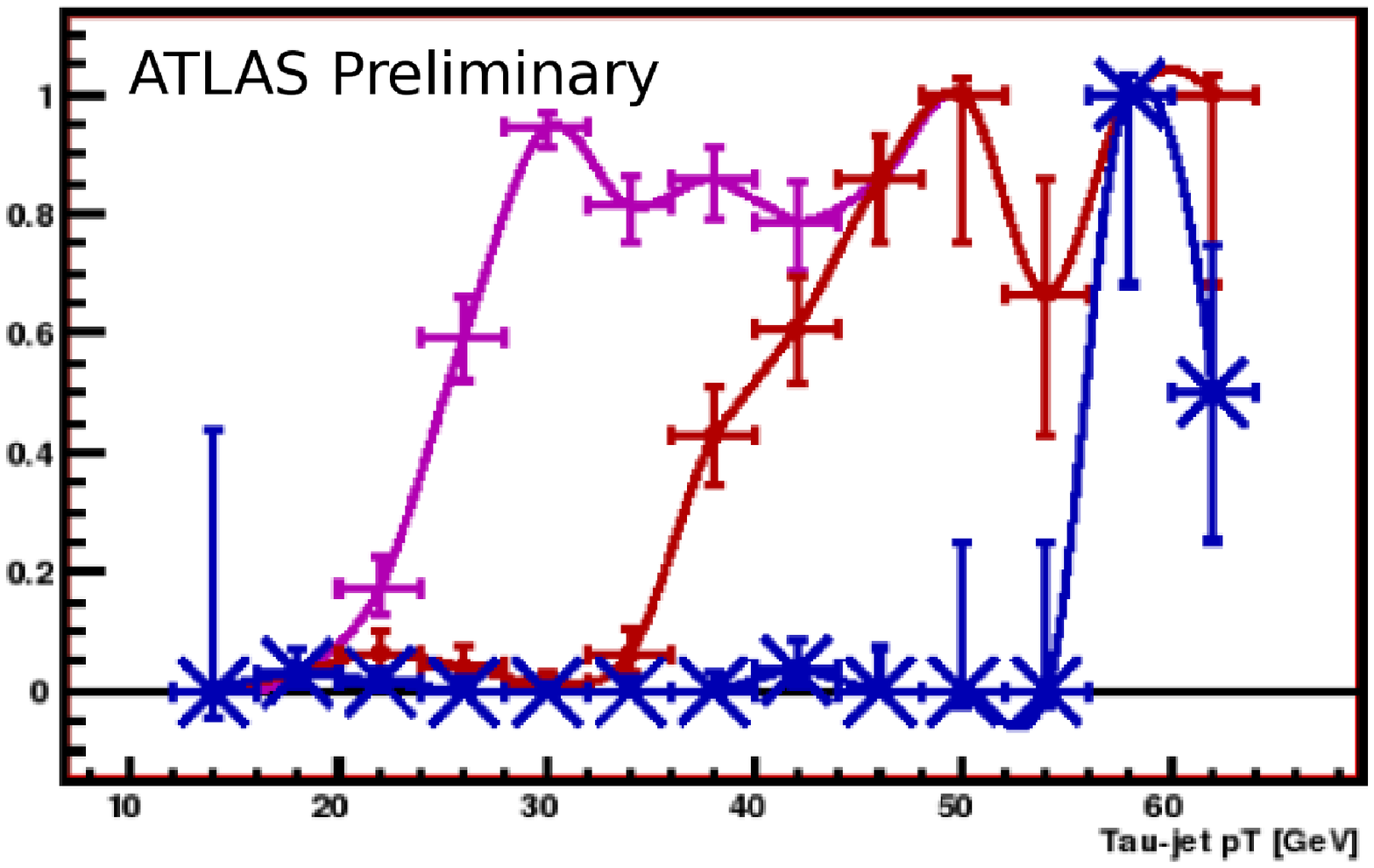}} \\
   \end{tabular}
  \caption{LVL1 tau trigger efficiency with respect to offline reconstructed taus estimated from the QCD jets events and compared with result
           for $\mathrm{W \rightarrow \tau \nu}$(left).
           Full trigger efficiency with respect to offline  reconstructed taus estimated from 
           from $\mathrm {Z \rightarrow \tau \tau \rightarrow \mu + hadrons + X}$  
           events for the tau\_25i, tau\_40 and tau\_60 trigger signatures (right).
            \label{EffEst}}       
  \end{center}
 \end{figure}
Figure~\ref{EffEst} (left) shows the tau20i LVL1 trigger signature efficiency with respect to the offline selection for the 
QCD jets selected by a minimum bias trigger and passing the tau offline selection. 
A comparison with the LVL1 trigger efficiency for $\mathrm {W \rightarrow \tau \nu}$ is also shown. 
Figure~\ref{EffEst} (right) shows the full trigger efficiency using a tag \& probe method. 
The Z sample is selected via single electron or muon triggers. After offline 
selection, the trigger lepton is used as a tag and the other side of the 
event is probed for a tau trigger candidate, and the efficiency of the tau 
trigger relative to offline reconstruction is measured. 
Both methods reproduce the shape and threshold of the 
trigger turn-on curves, but further work is needed to improve 
the background estimation and to gain a better understanding 
of the bias introduced by the event selection.
\section{CONCLUSIONS}
Many of the SM processes being investigated at ATLAS, as well as numerous BSM 
searches, contain tau leptons in their final states. Being able to 
trigger effectively on the tau leptons in these events will contribute 
to the success of the ATLAS experiment. 
The tau trigger algorithms and monitoring infrastructure are ready for the first data, 
and are being tested with the data collected with cosmic muons. 
The development of efficiency measurements methods using QCD and $\mathrm{Z\rightarrow \tau \tau}$ events is well advanced.

\end{document}